%% file: main.tex
\documentclass[bib]{statapress}

\usepackage[crop,newcenter,frame]{pagedims}

\usepackage{sj}

\usepackage{epsfig}
\usepackage{epstopdf}

\usepackage{stata}

\usepackage{shadow}

\usepackage{amsfonts}
\usepackage{amsmath}

\sjsetissue{$vv$}{$ii$}{$mm$}{$yyyy$}

\begin{document}

\input{stj}

\clearpage
\end{document}

%% file: stj.tex
\inserttype[st0001]{article}
\author{M. J. Grayling \& A. P. Mander}{%
  Michael J. Grayling\\Hub for Trials Methodology\\ Research, MRC Biostatistics Unit, \\Cambridge, UK\\mjg211@cam.ac.uk
  \and
  Adrian P. Mander\\Hub for Trials Methodology\\ Research, MRC Biostatistics Unit, \\Cambridge, UK\\adrian.mander@mrc-bsu.cam.ac.uk
}
\title[The multivariate normal and t distributions]{Calculations involving the multivariate normal and multivariate t distributions with and without truncation}
\maketitle

\begin{abstract}
This paper presents a set of Stata commands and Mata functions to evaluate different distributional quantities of the multivariate normal distribution, and a particular type of non-central multivariate \textit{t} distribution. Specifically, their densities, distribution functions, equicoordinate quantiles, and pseudo-random vectors can be computed efficiently, either in the absence or presence of variable truncation.
\end{abstract}

\input sj.tex

\bibliographystyle{sj}
\bibliography{mvn_references}

\begin{aboutauthors}
Michael J. Grayling is an Investigator Statistician at the Medical Research Council Biostatistics Unit in Cambridge, UK.

Adrian P. Mander is the Director of the Hub for Trial Methodology Research at the Medical Research Council Biostatistics Unit in Cambridge, UK.

\end{aboutauthors}

%% file: sj.tex
\section{Introduction}

The normal distribution plays an important role in statistics as much collected data are, or are assumed to be, normally distributed. Whilst the central limit theorem tells us that for any random sample $X_1,\dots,X_n$ of independent and identically distributed random variables with expected value $\mu$ and variance $\sigma^2$, $\sqrt{n}(\sum_{i=1}^nX_i/n-\mu)$ converges in distribution to $N(0,\sigma^2)$ as $n\rightarrow\infty$ \citep{Patel1996}. Moreover, the \textit{t} distribution, and its use in cases of normally distributed data of small sample size and of unknown variance, is also of considerable importance \citep{Ireland2010}. Consequently, functions such as \texttt{normal()} and \texttt{t()} are available for working with these distributions in Stata.

However, a considerable amount of statistical analysis performed is not univariate, but multivariate. As a result, the generalisation of the normal and \textit{t} distributions to higher dimensions; the multivariate normal (MVN) and multivariate \textit{t} distributions (MVT), are of increasing significance. Their broad use can be seen through introductions created for statisticians \citep{Tong2012,Kotz2004a}, natural scientists \citep{Bailer1997,Blaesild2002,Mazza2014,Feigelson2012} and social scientists \citep{Stevens2012,Howell2012}.

The complexity of these distributions makes computational analysis almost certainly necessary, and thus much research into their numerical analysis has been conducted \citep{Genz2009}. Many programming languages now have efficient code available for working with densities, evaluating distribution functions, finding quantiles, and generating pseudo-random variables (see, for example, \citet{Genz2014}). In Stata though, no code currently exists for working with the MVT distribution, and there are limitations to the support for the MVN distribution. Specifically, the \texttt{drawnorm} command allows users to generate pseudo-random samples from a MVN distribution, and \texttt{lnmvnormalden()} is available to evaluate its density. \citet{Cappallari2006} provided \texttt{mvnp} for the evaluation of cumulative MVN probabilities of any dimension through use of the Geweke-Hajivassiliou-Keane simulator \citep{Geweke1989,Hajivassiliou1998,Keane1994}. This simulator is also utilised in the Mata function \texttt{ghk()} \citep{Gates2006}. Whilst in Stata 15, \texttt{mvnormal()} and several similar mata commands were released to facilitate the computation of the MVN distribution function over any range of integration. However, to the best of our knowledge, no command is presently obtainable for easily determining equicoordinate quantiles of the MVN distribution.

Furthermore, no commands are currently available to perform calculations involving the MVN or MVT distribution, or their univariate counterparts, in the presence of variable truncation. In certain scenarios, it is not uncommon for one or more variables in an MVN or MVT setting to be bound to some particular finite interval. For example, the truncated normal distribution is utilised in modelling censored data through the Tobit model \citep{Tobin1958}.

In this work, we present several Stata commands and Mata functions for performing key calculations for any non-degenerate case of the MVN or MVT distribution, and for a particular type of non-central multivariate \textit{t} distribution (NCMVT), either in the presence or absence of variable truncation. We proceed by briefly summarising the algorithms underpinning these commands, before detailing their syntax, and finally demonstrating their use through several examples.

\section{Methods}

\subsection{Multivariate normal distribution}

We begin by considering a $k$-dimensional random variable $\boldsymbol{X}=(X_1,\dots,X_k)^\top$ with a (non-degenerate) MVN distribution. We denote this by $\boldsymbol{X} \sim N_k\left(\boldsymbol{\delta},\Sigma\right)$, where $\boldsymbol{\delta} \in \mathbb{R}^k$ is a location parameter, and $\Sigma \in \mathbb{M}^{k \times k}$ is a positive-definite covariance matrix. Many of the steps below revolve around the evaluation of $\Phi_k(\boldsymbol{a},\boldsymbol{b},\boldsymbol{\delta},\boldsymbol{\Sigma}) = \mathbb{P}(a_1\le X_1\le b_1,\dots,a_k\le X_k\le b_k)$, for $\boldsymbol{a}=(a_1,\dots,a_k)^\top$ and $\boldsymbol{b}=(b_1,\dots,b_k)^\top$ with $-\infty\le a_i<b_i\le\infty$ for $i=1,\dots,k$. Specifically
\begin{align*}
\Phi_k(\boldsymbol{a},\boldsymbol{b},\boldsymbol{\delta},\Sigma) &= \int_{a_1}^{b_1} \dots \int_{a_k}^{b_k} \phi_k(\boldsymbol{x},\boldsymbol{\delta},\Sigma)\ \mathrm{d}\boldsymbol{x},\\
&= \int_{a_1}^{b_1} \dots \int_{a_k}^{b_k} \frac{1}{\sqrt{(2\pi)^k|\Sigma|}}\text{exp}\left\{ -\frac{1}{2}(\boldsymbol{x}-\boldsymbol{\delta})^\top \Sigma^{-1} (\boldsymbol{x}-\boldsymbol{\delta}) \right\}\ \mathrm{d}\boldsymbol{x},
\end{align*}
for $\boldsymbol{x}=\left(x_1,\dots,x_k\right)^\top \in \mathbb{R}^k$, and where $|\Sigma|=\text{det}(\Sigma)$.

Several different methods are available for evaluating such integrals. For example, the mata function \texttt{mvnormalcv()} employs numerical quadrature for this task. In our code, we utilise a quasi-Monte Carlo integration algorithm over a randomised lattice after separation-of-variables has been performed (see \citet{Genz1992,Genz2002,Genz2009} for further details). Moreover, we employ variable re-ordering in order to improve efficiency as proposed by \citet{Gibson1994}. This is the approach recommended for such calculations by \citet{Genz2009}. The procedure is implemented in our Stata command \texttt{pmvnormal}. The algorithm employed requires specification of the number of shifts and the number of samples to use in the Monte Carlo integration. In addition, a value for the Monte Carlo confidence factor is necessitated. However, default values that have been shown to perform consistently well are provided, and thus little user alteration is required in practice. Increasing the number of shifts or samples theoretically brings about a reduction in the Monte Carlo error of the integration, but this comes at a cost to the run time of the commands.

In the above, $\phi_k \left(\boldsymbol{x},\boldsymbol{\delta},\Sigma\right)$ is the probability density function for $\boldsymbol{X}$. We later present a Stata command \texttt{mvnormalden} to directly compute the above density for any user specified choices of $\boldsymbol{x}$, $\boldsymbol{\delta}$ and $\Sigma$. 

Note that those with access to Stata 15 should utilise \texttt{mvnormalcv()} for evaluating $\Phi_k(\boldsymbol{a},\boldsymbol{b},\boldsymbol{\delta},\boldsymbol{\Sigma})$, as its execution time will be smaller. Additionally, as noted earlier, \texttt{lnmvnormalden()} could also be used for density calculations. The purpose of our commands is to provide a framework for these tasks with a consistent syntax to our commands performing previously unsupported derivations, and as a means of evaluating the MVN distribution function with earlier versions of Stata.

Also routinely of interest in multivariate analysis is the determination of equicoordinate quantiles of the MVN distribution. Determining the value, $q$, of the $(100)p$th equicoordinate quantile of $X$, requires us to solve the following equation
\[p = \Phi_k\{(-\infty,\dots,-\infty)^\top,(q,\dots,q)^\top,\boldsymbol{\delta},\Sigma\}.\]
Here, we achieve this in the Stata command \texttt{invmvnormal}, allowing either \texttt{mvnormalcv()} or the algorithm implemented in \texttt{pmvnormal} to be used to evaluate the right hand side of the above equation. Modified interval bisection is then utilised to search for the correct value of $q$.

Finally, in order to draw pseudo-random vectors with distribution $N_{k}(\boldsymbol{\delta},\Sigma)$, we utilise the fact that for $R$ with $RR^\top=\Sigma$, $\boldsymbol{x}=\boldsymbol{\delta}+R\boldsymbol{z}$ has the desired distribution when $\boldsymbol{z}=(z_1,\dots,z_k)^\top$ with $z_i\sim N(0,1)$. The $z_i$ can be generated using \texttt{rnormal()}. Several methods for generating such an $R$ are available, as discussed for example by \citet{Johnson1987}. Here we allow for eigen, singular value, and cholesky decomposition of $\Sigma$. Explicitly, our Stata command for this pseudo-random generation is \texttt{rmvnormal}. As for the density and distribution function evaluations discussed above, \texttt{drawnorm} provides similar functionality. In this case however, our command provides additional options as well as a consistent syntax.

\subsection{Multivariate $t$ distribution}

We now consider a $k$-dimensional random variable $\boldsymbol{X}=(X_1,\dots,X_k)^\top$ with a (non-degenerate) NCMVT distribution. We denote this by $\boldsymbol{X} \sim T_{k}\left(\boldsymbol{\delta},\Sigma,\nu\right)$, where $\boldsymbol{\delta}\in\mathbb{R}^k$ is now a vector of non-centrality parameters, $\Sigma \in \mathbb{M}^{k \times k}$ is a positive-definite scale matrix, and $\nu\in\mathbb{R}^+$ is a degrees of freedom parameter. There are in-fact several ways by which one can define an associated NCMVT integral. Here we consider what has been referred to as the location shifted NCMVT \citep{Genz2009}. See also \citet{Kotz2004b} for further details. This NCMVT distribution arises for example as the Bayesian posterior distribution for the regression coefficients in a linear regression \citep{Genz2014}. Note that in the case when $\boldsymbol{\delta}=\boldsymbol{0}$ the central MVT distribution is obtained.

Our integral of interest is as follows
\begin{small}
\begin{align*} \Psi_k(\boldsymbol{a},\boldsymbol{b},\boldsymbol{\delta},\Sigma,\nu)&=\int_{a_1}^{b_1} \dots \int_{a_k}^{b_k} \psi_k(\boldsymbol{x},\boldsymbol{\delta},\Sigma,\nu)\ \mathrm{d}\boldsymbol{x},\\
&= \int_{a_1}^{b_1} \dots \int_{a_k}^{b_k} \frac{\Gamma\left[ (\nu+k)/2 \right]}{\Gamma(\nu/2)\sqrt{(\nu\pi)^k|\Sigma|}} \left[ 1 + \frac{1}{\nu}(\boldsymbol{x} - \boldsymbol{\delta})^\top \Sigma^{-1} (\boldsymbol{x} - \boldsymbol{\delta}) \right]^{-(\nu + k)/2}\ \mathrm{d}\boldsymbol{x},
\end{align*}	
\end{small}

and can be evaluated using \texttt{mvt}. In addition, \texttt{mvtden} is available to evaluate the density $\psi_k(\boldsymbol{x},\boldsymbol{\delta},\Sigma,\nu)$.

We would again like to determine equicoordinate quantiles, this time solving the following equation
\[p = \Psi_k\{(-\infty,\dots,-\infty)^\top,(q,\dots,q)^\top,\boldsymbol{\delta},\Sigma,\nu\}.\]

Later, the stata command \texttt{invmvt} is described that performs this calculation. In both \texttt{mvt} and \texttt{invmvt}, the algorithm employed for numerical integration is a direct modification of that described above for the MVN distribution.

Finally, in order to draw pseudo-random vectors following a location shifted $T_{k}(\boldsymbol{\delta},\Sigma,\nu)$ distribution, we utilise the fact that if $\boldsymbol{y} \sim N_k(\boldsymbol{0},\Sigma)$ and $v \sim \chi_\nu^2$, then
\[ \boldsymbol{x} = \boldsymbol{\delta} + \boldsymbol{y}\sqrt{\frac{\nu}{v}}, \]
has the desired distribution. We draw pseudo-random vectors for $\boldsymbol{y}$ using the algorithm described earlier for the MVN distribution, with pseudo-random numbers $v$ acquired through \texttt{rchi2()}. The Stata command for this is \texttt{rmvt}.

\subsection{Truncated multivariate normal and multivariate $t$ distributions}

We additionally consider truncated versions of the MVN and NCMVT distributions discussed above. We denote these by $\boldsymbol{X} \sim TN_{k}\left(\boldsymbol{\delta},\Sigma,\boldsymbol{l},\boldsymbol{u}\right)$ and $\boldsymbol{X} \sim TT_{k}\left(\boldsymbol{\delta},\Sigma,\nu,\boldsymbol{l},\boldsymbol{u}\right)$ respectively. Here, $\boldsymbol{l}=(l_1,\dots,l_k)^\top\in\mathbb{R}^k$ and $\boldsymbol{u}=(u_1,\dots,u_k)^\top\in\mathbb{R}^k$, with $l_i<u_i$ for $i=1,\dots,k$, are the lower and upper truncation points for $\boldsymbol{X}$. That is, we restrict such that $l_i \le X_i \le u_i$ for $i=1,\dots,k$. The distribution functions for these variables are

\begin{align*}
\Phi_{k}'(\boldsymbol{a},\boldsymbol{b},\boldsymbol{\delta},\Sigma,\boldsymbol{l},\boldsymbol{u}) &= \int_{\max(a_1,l_1)}^{\min(b_1,u_1)} \dots \int_{\max(a_k,l_k)}^{\min(b_k,u_k)} \phi_k'(\boldsymbol{x},\boldsymbol{\delta},\Sigma,\boldsymbol{l},\boldsymbol{u})\ \mathrm{d}\boldsymbol{x},\\
&= \int_{\max(a_1,l_1)}^{\min(b_1,u_1)} \dots \int_{\max(a_k,l_k)}^{\min(b_k,u_k)} \frac{\phi_k(\boldsymbol{x},\boldsymbol{\delta},\Sigma)}{\Phi_{k}(\boldsymbol{l},\boldsymbol{u},\boldsymbol{\delta},\Sigma)}\ \mathrm{d}\boldsymbol{x},\\
\Psi_{k}'(\boldsymbol{a},\boldsymbol{b},\boldsymbol{\delta},\Sigma,\nu,\boldsymbol{l},\boldsymbol{u})&=\int_{\max(a_1,l_1)}^{\min(b_1,u_1)} \dots \int_{\max(a_k,l_k)}^{\min(b_k,u_k)} \psi_k'(\boldsymbol{x},\boldsymbol{\delta},\Sigma,\nu,\boldsymbol{l},\boldsymbol{u})\ \mathrm{d}\boldsymbol{x},\\
&=\int_{\max(a_1,l_1)}^{\min(b_1,u_1)} \dots \int_{\max(a_k,l_k)}^{\min(b_k,u_k)} \frac{\psi_k(\boldsymbol{x},\boldsymbol{\delta},\Sigma,\nu)}{\Psi_{k}(\boldsymbol{l},\boldsymbol{u},\boldsymbol{\delta},\Sigma,\nu)}\ \mathrm{d}\boldsymbol{x},
\end{align*}

with $\phi_k'(\boldsymbol{x},\boldsymbol{\delta},\Sigma,\boldsymbol{l},\boldsymbol{u})$ and $\psi_k'(\boldsymbol{x},\boldsymbol{\delta},\Sigma,\nu,\boldsymbol{l},\boldsymbol{u})$ their respective densities. Our Stata commands to evaluate these quantities are \texttt{tmvnormal}, \texttt{tmvt}, \texttt{tmvnormalden}, and \texttt{tmvtden}.

We further provide commands \texttt{invtmvnormal} and \texttt{invtmvt} to determine the values of $q$ such that
\[p = \Phi_k'\{(-\infty,\dots,-\infty)^\top,(q,\dots,q)^\top,\boldsymbol{\delta},\Sigma,\boldsymbol{l},\boldsymbol{u}\},\]
or
\[p = \Psi_k'\{(-\infty,\dots,-\infty)^\top,(q,\dots,q)^\top,\boldsymbol{\delta},\Sigma,\nu,\boldsymbol{l},\boldsymbol{u}\}.\]

Finally, we use rejection sampling in the commands \texttt{rtmvnormal} and \texttt{rtmvt} to generate random variables with a $ TN_{k}\left(\boldsymbol{\delta},\Sigma,\boldsymbol{l},\boldsymbol{u}\right)$ or $TT_{k}\left(\boldsymbol{\delta},\Sigma,\nu,\boldsymbol{l},\boldsymbol{u}\right)$ distribution. For example, in \texttt{rtmvnormal}, a random draw from the $N_{k}\left(\boldsymbol{\delta},\Sigma\right)$ is made. If $l_i \le X_i \le u_i$ for $i=1,\dots,k$ then this draw is retained, otherwise it is rejected and the process is repeated.

\section{Syntax}

Note that all of the Stata commands presented here are written as .ado files, utilising Mata for computational efficiency wherever possible. This allows all desired calculations to be performed easily from within Stata. Corresponding Mata functions for each of the Stata commands are provided for convenience given the matrix based nature of this work.

\subsection{Stata commands}

\begin{stsyntax}
mvnormalden, x(numlist) \underbar{me}an(numlist) \underbar{s}igma(string) \optional{\underbar{log}density}
\end{stsyntax}

\begin{stsyntax}
pmvnormal, \underbar{low}er(numlist miss) \underbar{upp}er(numlist miss) \underbar{me}an(numlist) \underbar{s}igma(string)
\optional{\underbar{shi}fts(integer 12) \underbar{sam}ples(integer 1000) \underbar{alp}ha(real 3)}
\end{stsyntax}

\begin{stsyntax}
invmvnormal, p(real) \underbar{me}an(numlist) \underbar{s}igma(string)
\optional{\underbar{t}ail(string) \underbar{it}ermax(integer 1000000) \underbar{tol}erance(real 0.000001) \underbar{int}egrator(string) \underbar{shi}fts(integer 12) \underbar{sam}ples(integer 1000)}
\end{stsyntax}

\begin{stsyntax}
	rmvnormal, \underbar{me}an(numlist) \underbar{s}igma(string) \optional{n(integer 1) \underbar{meth}od(string)}
\end{stsyntax}

\begin{stsyntax}
mvtden, x(numlist) \underbar{del}ta(numlist) \underbar{s}igma(string) \optional{df(real 1) \underbar{log}density}
\end{stsyntax}

\begin{stsyntax}
mvt, \underbar{low}er(numlist miss) \underbar{upp}er(numlist miss) \underbar{del}ta(numlist) \underbar{s}igma(string)
\optional{df(real 1) \underbar{shi}fts(integer 12) \underbar{sam}ples(integer 1000) \underbar{alp}ha(real 3)}
\end{stsyntax}

\begin{stsyntax}
invmvt, p(real) \underbar{del}ta(numlist) \underbar{s}igma(string)
\optional{df(real 1) \underbar{t}ail(string) \underbar{it}ermax(integer 1000000) \underbar{tol}erance(real 0.000001) \underbar{shi}fts(integer 12) \underbar{sam}ples(integer 1000)}
\end{stsyntax}

\begin{stsyntax}
rmvt, \underbar{del}ta(numlist) \underbar{s}igma(string) \optional{df(real 1) n(integer 1) \underbar{meth}od(string)}
\end{stsyntax}

\begin{stsyntax}
	tmvnormalden, x(numlist) \underbar{me}an(numlist) \underbar{s}igma(string) \underbar{lowert}runcation(numlist miss) \underbar{uppert}runcation(numlist miss) \optional{\underbar{log}density \underbar{int}egrator(string) \underbar{shi}fts(integer 12) \underbar{sam}ples(integer 1000)}
\end{stsyntax}

\begin{stsyntax}
	tmvnormal, \underbar{low}er(numlist miss) \underbar{upp}er(numlist miss) \underbar{me}an(numlist) \underbar{s}igma(string)
	\underbar{lowert}runcation(numlist miss) \underbar{uppert}runcation(numlist miss) \optional{\underbar{int}egrator(string) \underbar{shi}fts(integer 12) \underbar{sam}ples(integer 1000)}
\end{stsyntax}

\begin{stsyntax}
	invtmvnormal, p(real) \underbar{me}an(numlist) \underbar{s}igma(string) \underbar{lowert}runcation(numlist miss) \underbar{uppert}runcation(numlist miss)
	\optional{\underbar{t}ail(string) \underbar{it}ermax(integer 1000000) \underbar{tol}erance(real 0.000001) \underbar{int}egrator(string) \underbar{shi}fts(integer 12) \underbar{sam}ples(integer 1000)}
\end{stsyntax}

\begin{stsyntax}
	rtmvnormal, \underbar{me}an(numlist) \underbar{s}igma(string) \underbar{lowert}runcation(numlist miss) \underbar{uppert}runcation(numlist miss) \optional{n(integer 1) \underbar{meth}od(string)}
\end{stsyntax}

\begin{stsyntax}
	tmvtden, x(numlist) \underbar{del}ta(numlist) \underbar{s}igma(string) \underbar{lowert}runcation(numlist miss) \underbar{uppert}runcation(numlist miss) \optional{df(real 1) \underbar{log}density \underbar{shi}fts(integer 12) \underbar{sam}ples(integer 1000)}
\end{stsyntax}

\begin{stsyntax}
	tmvt, \underbar{low}er(numlist miss) \underbar{upp}er(numlist miss) \underbar{del}ta(numlist) \underbar{s}igma(string) \underbar{lowert}runcation(numlist miss) \underbar{uppert}runcation(numlist miss)
	\optional{df(real 1) \underbar{shi}fts(integer 12) \underbar{sam}ples(integer 1000)}
\end{stsyntax}

\begin{stsyntax}
	invtmvt, p(real) \underbar{del}ta(numlist) \underbar{s}igma(string) \underbar{lowert}runcation(numlist miss) \underbar{uppert}runcation(numlist miss)
	\optional{df(real 1) \underbar{t}ail(string) \underbar{it}ermax(integer 1000000) \underbar{tol}erance(real 0.000001) \underbar{shi}fts(integer 12) \underbar{sam}ples(integer 1000)}
\end{stsyntax}

\begin{stsyntax}
	rtmvt, \underbar{del}ta(numlist) \underbar{s}igma(string) \underbar{lowert}runcation(numlist miss) \underbar{uppert}runcation(numlist miss) \optional{df(real 1) n(integer 1) \underbar{meth}od(string)}
\end{stsyntax}

Here, the above options are as follows

\hangpara{\texttt{x} is the vector of quantiles at which a density is sought (see $\boldsymbol{x}$ above).}

\hangpara{\texttt{\underbar{me}an} is the location parameter $\boldsymbol{\delta}$ of a MVN or truncated MVN distribution. In the case of a MVN distribution, this is the expected value of the distribution.}

\hangpara{\texttt{\underbar{s}igma} is the matrix $\Sigma$ described above. This must be symmetric positive-definite. In the case of a MVN distribution this is the variables covariance matrix. This is not the case for the NCMVT distribution, or for truncated distributions.}

\hangpara{\texttt{\underbar{log}density} specifies that the log of the evaluated density should be returned.}

\hangpara{\texttt{\underbar{low}er} is the vector of lower limits, denoted by $\boldsymbol{a}$ above. Use \texttt{.} to indicate a value is $-\infty$.}

\hangpara{\texttt{\underbar{upp}er} is the vector of upper limits, denoted by $\boldsymbol{b}$ above. Use \texttt{.} to indicate a value is $+\infty$.}

\hangpara{\texttt{\underbar{shi}fts} is the number of shifts of the Quasi-Monte Carlo integration algorithm to use. It must be a strictly positive integer. In \texttt{invmvnormal} and \texttt{invtmvnormal} this will only have an effect if \texttt{integrator} is set to \texttt{pmvnormal}.}

\hangpara{\texttt{\underbar{sam}ples} is the number of samples in each shift of the Quasi-Monte Carlo integration algorithm to use. It must be a strictly positive integer. As above, in \texttt{invmvnormal} and \texttt{invtmvnormal} this will only have an effect if \texttt{integrator} is set to \texttt{pmvnormal}.}

\hangpara{\texttt{\underbar{alp}ha} is the value of the Monte Carlo confidence factor to use in estimating the error in the returned value of the integral when using \texttt{pmvnormal} or \texttt{mvt}. It must be strictly positive.}

\hangpara{\texttt{p} is a desired probability, denoted by $p$ above. It must be strictly between 0 and 1.}

\hangpara{\texttt{\underbar{t}ail} specifies which quantile should be computed. For \texttt{invmvnormal} \texttt{"lower"} gives the $q$ such that
\[p = \Phi_k\{(-\infty,\dots,-\infty)^\top,(q,\dots,q)^\top,\boldsymbol{\delta},\Sigma\},\]
\texttt{"upper"} such that
\[p=\Phi_k\{(q,\dots,q)^\top,(\infty,\dots,\infty)^\top,\boldsymbol{\delta},\Sigma\},\]
and \texttt{"both"} such that
\[p=\Phi_k\{(-q,\dots,-q)^\top,(q,\dots,q)^\top,\boldsymbol{\delta},\Sigma\}.\]
Analogous statements hold for \texttt{invmvt}, \texttt{invtmvnormal}, and \texttt{invtmvt}.}

\hangpara{\texttt{\underbar{it}ermax} specifies the maximum allowed number of iterations in the implemented modified interval bisection algorithm.}

\hangpara{\texttt{\underbar{tol}erance} specifies the desired tolerance for termination of the implemented modified interval bisection algorithm.}

\hangpara{\texttt{n} is the number of random vectors to generate from a chosen distribution. It must be a strictly positive integer.}

\hangpara{\texttt{\underbar{del}ta} is the vector of non-centrality parameters $\boldsymbol{\delta}$ of a NCMVT or truncated NCMVT distribution. Note that this may not be these distributions expected value.}

\hangpara{\texttt{\underbar{lowert}runcation} is the vector of lower truncation limits, denoted by $\boldsymbol{l}$ above. Use \texttt{.} to indicate a value is $-\infty$.

\hangpara{\texttt{\underbar{uppert}runcation} is the vector of upper truncation limits, denoted $\boldsymbol{u}$ above. Use \texttt{.} to indicate a value is $+\infty$.

\hangpara{\texttt{df} is the degrees of freedom parameter, $\nu$, of a NCMVT or truncated NCMVT distribution.}

\subsection{Mata functions}

Mata functions are provided with a naming convention that adds the suffix `\texttt{\_mata}' on to the command names above. All inputs are listed in the same order as the Stata commands, and have the exact same name and interpretation. Input types are specified in the obvious manner. That is, \texttt{delta} and \texttt{lower} must be real vectors, \texttt{df} a real scalar, \texttt{method} a string, and similarly for other inputs. For this reason, we omit their specification here. However, as an example

\begin{stsyntax}
	mvtden\_mata(real vector x, real vector delta, real matrix Sigma, real scalar df, string logdensity)
\end{stsyntax}

\section{Examples}

\subsection{Density comparison}

In order to demonstrate the use of \texttt{mvnormalden}, \texttt{mvtden}, \texttt{tmvnormalden}, and \texttt{tmvtden}, we plot and compare densities in the case with
\[ \boldsymbol{\delta}=\begin{pmatrix} 0 \\ 0 \end{pmatrix}, \quad \Sigma = \begin{pmatrix} 1 & 0.5 \\ 0.5 & 1 \end{pmatrix}, \quad \nu=1, \quad \boldsymbol{l}=\begin{pmatrix} -1.5 \\ -1.5 \end{pmatrix}, \quad \boldsymbol{u}=\begin{pmatrix} 1.5 \\ 1.5 \end{pmatrix}.\]

We first loop over a grid of values in the region $[-3,3]^2$. At each point we compute the density of a MVN, MVT, truncated MVN, and truncated MVT distribution with the above parameters

\begin{stlog}
	\input{Example1part1.log.tex}
\end{stlog}

We then pass our results to \texttt{twoway contour} to produce Figure 1

\begin{stlog}
	\input{Example1part2.log.tex}
\end{stlog}

\begin{figure}[hbtp]
\centering
\label{fig1}
\includegraphics[width=0.9\textwidth]{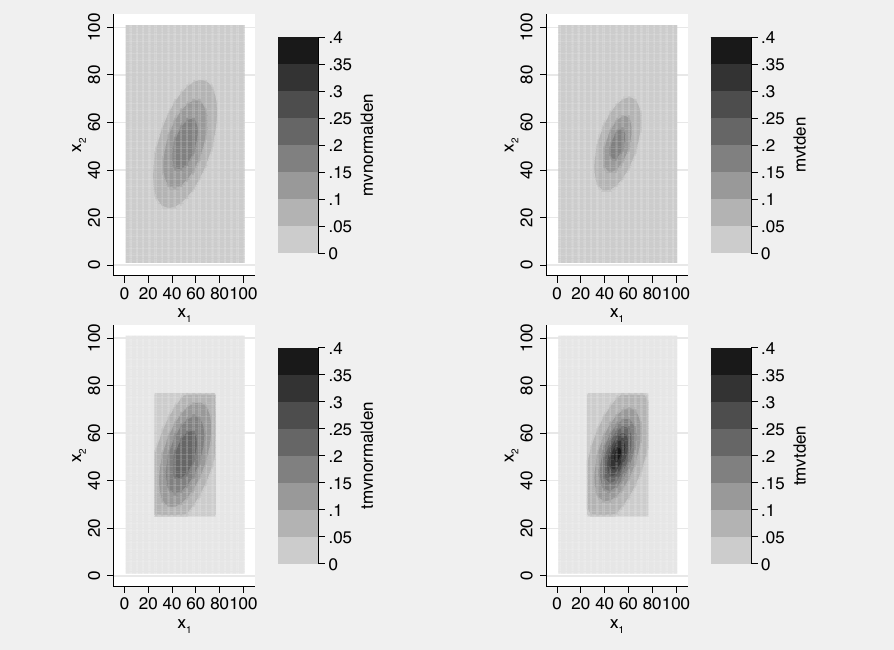}
\caption{Densities of a multivariate normal (mvnormalden), multivariate \textit{t} (mvtden), truncated multivariate normal (tmvnormalden), and truncated multivariate \textit{t} distribution (tmvtden), are shown.}
\end{figure}

As would be expected, we observe that the density of the MVT distribution with $\nu=1$ degree of freedom is more homogeneous than that of the MVN distribution. This explains why tail probabilities are larger for MVT distributions with low degrees of freedom. Truncation increases the density at each point in the region $[-1.5,1.5]^2$ compared to the corresponding non-truncated variable.

\subsection{Goodness of fit tests for multivariate normality}

Stata provides, through \texttt{mvtest normality}, a set of statistical tests of multivariate normality. Here, we explore the results of applying these tests to random variables drawn using \texttt{rmvt} and \texttt{rtmvt}. Specifically, we consider a bivariate MVT distribution with
\[ \boldsymbol{\delta}=\begin{pmatrix} 0 \\ 0 \end{pmatrix}, \quad \Sigma = \begin{pmatrix} 1 & 0.5 \\ 0.5 & 1 \end{pmatrix}.\]
We also examine results for a variable that is a mixture of two truncated MVT distributions, each with the above value for $\Sigma$, but one with
\[ \boldsymbol{\delta}=\begin{pmatrix} -2 \\ -2 \end{pmatrix}, \quad \boldsymbol{l}=\begin{pmatrix} -2.5 \\ -2.5 \\ \end{pmatrix}, \quad \boldsymbol{u}=\begin{pmatrix} -1.5 \\ -1.5 \end{pmatrix},\]
and the other
\[ \boldsymbol{\delta}=\begin{pmatrix} 2 \\ 2 \end{pmatrix}, \quad \boldsymbol{l}=\begin{pmatrix} 1.5 \\ 1.5 \\ \end{pmatrix}, \quad \boldsymbol{u}=\begin{pmatrix} 2.5 \\ 2.5 \end{pmatrix}.\]
We additionally consider several possible values for $\nu$. We draw $n$ samples from the MVT distribution, for multiple values of $n$, 1000 times in order to compute an average $p$-value from the Doornik-Hansen omnibus test. This process is also performed for the truncated MVT distributions, except $n/2$ samples are drawn from each and then combined in to a single dataset, so that the overall sample size is retained at $n$. We achieve this with the following code

\begin{stlog}
	\input{Example2.log.tex}
\end{stlog}

Observe that we have utilised cholesky decomposition of $\Sigma$ in generating these random variables. For most distributions the choice of decomposition will have little effect on the speed and quality of the generated random variables. However, when the dimension $k$ is large, this choice should be made more carefully.

Now, the results of the Doornik-Hansen tests are summarised in Table 1. We can see that, as would be expected, increasing the value of $\nu$ for the MVT distribution typically increases the average $p$-value for each value of $n$, as the resultant distribution is closer to that of a corresponding MVN distribution. Whilst the pattern is slightly less clear as we increase $n$, in general this reduces the average $p$-value, as more data allows the test to more precisely categorise the distribution of the generated variables.

For the datasets which are a mixture of the truncated NCMVT variables, perhaps surprisingly the average $p$-values are often larger than the corresponding value for the MVT distribution. In no instance is the Doornik-Hansen omnibus test significant on average at the 5\% level. That is, in no case does it on average correctly reject the null hypothesis of multivariate normality. This is likely a consequence of the chosen values for $\boldsymbol{\delta}$, $\boldsymbol{l}$, and $\boldsymbol{u}$ in the two truncated MVT distributions. When the two simulated datasets are combined, they likely resemble a MVN distribution with $\boldsymbol{\delta}=\boldsymbol{0}$.

\def\arraystretch{1.2}
\begin{table}[htb]
	\begin{center}
		\caption{Average $p$-values from the Doornik-Hansen omnibus test across 1000 replicate generated datasets are shown, as a function of the sample size $n$ and the degrees of freedom $\nu$. Results are displayed for a particular multivariate $t$ distribution, and a variable that is a mixture of two truncated multivariate $t$ distributions, as described in the text.}
		\label{tab1}
		\begin{tabular}{rrrrrr}
			\hline
			\multicolumn{6}{c}{Mixture of multivariate $t$ distributions}\\
			\hline
			& $\nu=2$ & $\nu=5$ & $\nu=10$ & $\nu=50$ & $\nu=100$ \\
			\hline
			$n=10$  & 0.2948 & 0.4069 & 0.4540 & 0.4851 & 0.4882 \\
			$n=25$  & 0.1040 & 0.3200 & 0.4217 & 0.5066 & 0.5109 \\
			$n=50$  & 0.0063 & 0.1502 & 0.3281 & 0.5065 & 0.5090 \\
			$n=100$ & 0.0003 & 0.0574 & 0.2326 & 0.4656 & 0.5046 \\
			\hline
			\multicolumn{6}{c}{Mixture of truncated non-central multivariate t distributions}\\
			\hline
			& $\nu=2$ & $\nu=5$ & $\nu=10$ & $\nu=50$ & $\nu=100$ \\
			\hline
			$n=10$  & 0.6612 & 0.6502 & 0.6398 & 0.6402 & 0.6483 \\
			$n=25$  & 0.6634 & 0.6466 & 0.6377 & 0.6138 & 0.6053 \\
			$n=50$  & 0.4281 & 0.4581 & 0.4720 & 0.4712 & 0.4446 \\
			$n=100$ & 0.1439 & 0.1942 & 0.2300 & 0.2443 & 0.2406 \\
		\end{tabular}
	\end{center}
\end{table}

\subsection{Familywise error-rate using Bonferroni and Dunnett corrections}

As a brief example of the usage of \texttt{pmvnormal} and \texttt{invmvnormal}, consider a statistical test of the following hypotheses
\[ H_{0i} : \mu_i \le 0, \quad H_{1i} : \mu_i > 0,\quad i=1,2,3. \]
Suppose our tests are based on test statistics $Z_i$, $i=1,2,3$, and it is known that
\[ (Z_1,Z_2,Z_3)^\top \sim N_3\left\{\begin{pmatrix} \mu_1 \\ \mu_2 \\ \mu_3 \end{pmatrix},\begin{pmatrix} 1 & 0.5 & 0.5 \\ 0.5 & 1 & 0.5 \\ 0.5 & 0.5 & 1 \end{pmatrix}\right\}. \]
The familywise error-rate, the probability of incorrectly rejecting one or more of the null hypotheses, can be controlled to a level $\alpha$ using the Bonferroni correction by rejecting $H_{0i}$ if $Z_i>\Phi^{-1}_1(1-\alpha/3)$. The Dunnett correction in comparison rejects if $Z_i>r$, where $r$ is the solution to
\[1-\alpha = \Phi_3\left\{(-\infty,\dots,-\infty)^\top,(r,\dots,r)^\top,\begin{pmatrix} 0 \\ 0 \\ 0 \end{pmatrix},\begin{pmatrix} 1 & 0.5 & 0.5 \\ 0.5 & 1 & 0.5 \\ 0.5 & 0.5 & 1 \end{pmatrix}\right\}.\]
We can use \texttt{invmvnormal} to determine this value of $r$, with the following code

\begin{stlog}
	\input{Example3part1.log.tex}
\end{stlog}

Returned along with the quantile, which is determined to be approximately 2.06, is an estimate of the error in this value. Also provided is a flag variable, which takes the value 0 if the interval bisection algorithm converged with no issues. A value of the objective function (\texttt{fquantile}) we attempt to find the root of, and the number of iterations required for convergence, are additionally listed.

We can then verify that this value of $r$ will control the familywise error-rate to $\alpha$ using \texttt{pmvnormal} as follows, additionally evaluating the true familywise error-rate when using the Bonferroni correction

\begin{stlog}
	\input{Example3part2.log.tex}
\end{stlog}

We can now see the conservatism of the Bonferroni correction: the true familywise error-rate is approximately 4.3\%.

\subsection{Orthont probabilities in the presence of truncation}

The value of $\Phi_k\{(0,\dots,0)^\top,(\infty,\dots,\infty)^\top,\boldsymbol{\delta},\boldsymbol{\Sigma}\}$ is often referred to as an orthont probability. Here, as a final example, we examine how $\Phi_k'\{(0,\dots,0)^\top,(\infty,\dots,\infty)^\top,\boldsymbol{\delta},\boldsymbol{\Sigma},(t,\dots,t)^\top,(\infty,\dots,\infty)^\top\}$ changes in $t$, in order to demonstrate the usage of \texttt{tmvnormal}.

We consider again the case with 

\[ \boldsymbol{\delta}=\begin{pmatrix} 0 \\ 0 \\ 0 \end{pmatrix}, \quad \Sigma = \begin{pmatrix} 1 & 0.5 & 0.5 \\ 0.5 & 1 & 0.5 \\ 0.5 & 0.5 & 1 \end{pmatrix}.\]

Using \texttt{pmvnormal}, we can establish that $\Phi_3\{(0,\dots,0)^\top,(\infty,\dots,\infty)^\top,\boldsymbol{\delta},\boldsymbol{\Sigma}\}\approx0.25$

\begin{stlog}
	\input{Example4part1.log.tex}
\end{stlog}

The second value returned in the above is, as in Section 4.3, an estimate of the error in the integral. As we can see, \texttt{pmvnormal} is again easily able to control this error to a small level.

Then, as stated, we can evaluate how this probability changes in $t$ for a truncated MVN distribution as follows

\begin{stlog}
	\input{Example4part2.log.tex}
\end{stlog}

Here, \texttt{twoway line} is used to produce Figure 2. As expected, when $t$ is highly negative, the value of $\Phi_3'\{(0,\dots,0)^\top,(\infty,\dots,\infty)^\top,\boldsymbol{\delta},\boldsymbol{\Sigma},(t,\dots,t)^\top,(\infty,\dots,\infty)^\top\}$ is approximately equal to that of the corresponding non-truncated distribution. This value increases up to a value of one when $t\ge0$, since at this point truncation implies we must have that $X_i\ge0$ for $i=1,2,3$.

\begin{figure}[hbtp]
	\centering
	\label{fig2}
	\includegraphics[width=0.9\textwidth]{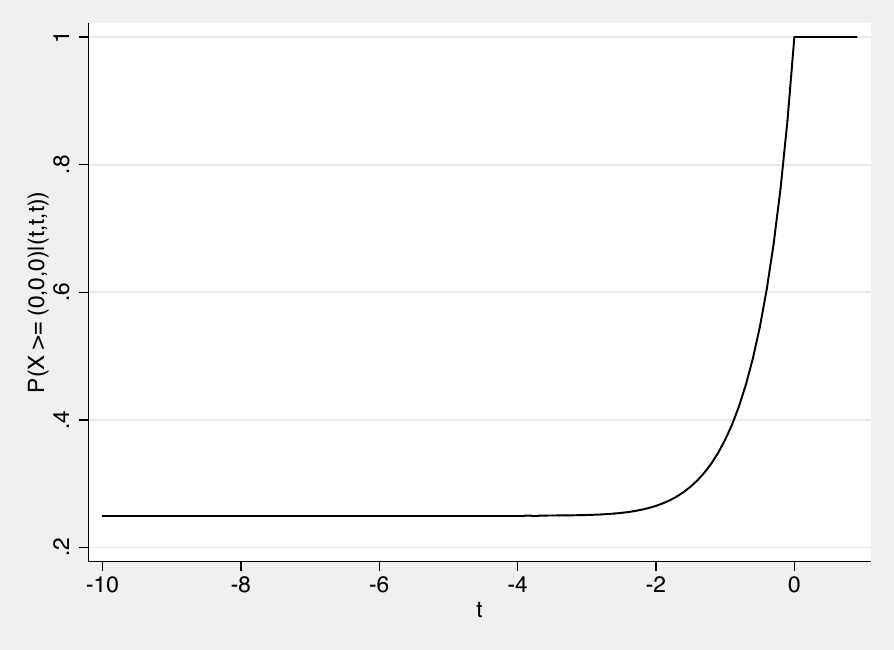}
	\caption{$\Phi_3'\{(t,\dots,t)^\top\}\equiv\Phi_3'\{(0,\dots,0)^\top,(\infty,\dots,\infty)^\top,\boldsymbol{\delta},\boldsymbol{\Sigma},(t,\dots,t)^\top,(\infty,\dots,\infty)^\top\}$ is shown as a function of $t$ for a particular truncated multivariate normal distribution.}
\end{figure}

\section{Conclusion}

The MVN and MVT distributions are extremely important in a wide range of statistical problems faced by researchers in many fields. Here we have extended Stata users ability to work with these key distributions; presenting several commands to allow densities, pseudo-random draws, distribution functions, and equicoordinate quantiles to be computed quickly and efficiently, both with and without variable truncation. In particular, using algorithms developed by \citet{Genz2009}, our programs require little input from the user to determine probabilities the distributions fall in any range of integration. Whilst this is possible from the MVN distribution using \texttt{mvnormalcv()}, our commands are the first for the NCMVT distribution, as well as for truncated MVN and NCMVT distributions.

However, whilst the algorithms utilised here for the computation of probabilities over arbitrary ranges of integration are efficient, there are many alternatives available. Moreover, we have considered only one possible definition of a NCMVT distribution. Another important definition, used in the computation of the power of multiple contrast tests under a normality assumption \citep{Genz2014}, was proposed by Kshirsagar \citet{Kotz2004b}. Thus, further development of the functions presented here will seek to make additional algorithms available as an option in the distribution function commands, as well as to include an option to change between the shifted NCMVT considered here and that of Kshirsagar available in the NCMVT and truncated NCMVT commands.

\section{Acknowledgements}

Michael J. Grayling is supported by the Wellcome Trust (Grant Number 099770/Z/12/Z). Adrian P. Mander is supported by the Medical Research Council (Grant Number MC\_UP\_1302/2).

%% file: Example1part1.log.tex
{\smallskip}
. mat Sigma        = J(2, 2, 0.5) + 0.5*I(2)
{\smallskip}
. mat mvnormalden  = J(101, 101, 0)
{\smallskip}
. mat mvtden       = J(101, 101, 0)
{\smallskip}
. mat tmvnormalden = J(101, 101, 0)
{\smallskip}
. mat tmvtden      = J(101, 101, 0)
{\smallskip}
. local i = 1
{\smallskip}
. foreach x1 of numlist -3(0.06)3 {\lbr}
  2.   local j = 1
  3.   foreach x2 of numlist -3(0.06)3 {\lbr}
  4.     qui mvnormalden, x(`x1', `x2') mean(0, 0) sigma(Sigma)
  5.     mat mvnormalden[`i', `j'] = r(density)
  6.     qui mvtden, x(`x1', `x2') delta(0, 0) sigma(Sigma) df(`nu')
  7.     mat mvtden[`i', `j'] = r(density)
  8.     qui tmvnormalden, x(`x1', `x2') mean(0, 0) sigma(Sigma) lowert(-1.5, -1.5) 
 > uppert(1.5, 1.5)
  9.     mat tmvnormalden[`i', `j'] = r(density)
 10.     qui tmvtden, x(`x1', `x2') delta(0, 0) sigma(Sigma) lowert(-1.5, -1.5)
 > uppert(1.5, 1.5) df(1)
 11.     mat tmvtden[`i', `j'] = r(density)
 12.     local `j++'
 13.   {\rbr}
 14.   local `i++'
 15. {\rbr}

%% file: Example1part2.log.tex
. qui svmat mvnormalden
{\smallskip}
. qui svmat mvtden
{\smallskip}
. qui svmat tmvnormalden
{\smallskip}
. qui svmat tmvtden
{\smallskip}
. gen row = _n
{\smallskip}
. qui reshape long mvnormalden mvtden tmvnormalden tmvtden, i(row) j(col)
{\smallskip}
. local opt "aspect(2) ccuts(0 0.05 0.1 0.15 0.2 0.25 0.3 0.35 0.4) 
> xtitle("x{\lbr}subscript:1{\rbr}") ytitle("x{\lbr}subscript:2{\rbr}")"
{\smallskip}
. twoway contour mvnormalden row col, nodraw saving(g1, replace) `opt'
(file g1.gph saved)
{\smallskip}
. twoway contour mvtden row col, nodraw saving(g2, replace) `opt'
(file g2.gph saved)
{\smallskip}
. twoway contour tmvnormalden row col, nodraw saving(g3, replace) `opt'
(file g3.gph saved)
{\smallskip}
. twoway contour tmvtden row col, nodraw saving(g4, replace) `opt'
(file g4.gph saved)
{\smallskip}
. graph combine g1.gph g2.gph g3.gph g4.gph, common imargin(0 0 0 0) rows(2)
> cols(2) scheme(sj)

%% file: Example2.log.tex
. set seed 2
{\smallskip}
. mat Sigma = J(2, 2, 0.5) + 0.5*I(2)
{\smallskip}
. mat list Sigma
{\smallskip}
. mat gof_pvaluesum_mvn = J(4, 5, 0)
{\smallskip}
. mat gof_pvaluesum_mix = J(4, 5, 0)
{\smallskip}
. local i = 1
{\smallskip}
. foreach n of numlist 10 25 50 100 {\lbr}
  2.   local j = 1
  3.   foreach df of numlist 2 5 10 50 100 {\lbr}
  4.     foreach rep of numlist 1/1000 {\lbr}
  5.       rmvt, delta(0, 0) sigma(Sigma) df(`df') n(`n') method(chol)
  6.       mat randsamp = r(rmvt)
  7.       qui svmat randsamp
  8.       qui mvtest normality randsamp1 randsamp2
  9.       mat gof_pvaluesum_mvn[`i', `j'] = gof_pvaluesum_mvn[`i', `j'] + r(p_dh)
 10.       drop randsamp1 randsamp2
 11.       rtmvt, delta(-2, -2) sigma(Sigma) df(`df') n(`n') method(chol)
 > lowert(-2.5, -2.5) uppert(-1.5, -1.5)
 12.       mat randsamp = r(rtmvt)
 13.       rtmvt, delta(2, 2) sigma(Sigma) df(`df') n(`n') method(chol)
 > lowert(1.5, 1.5) uppert(2.5, 2.5)
 14.       mat randsamp = (randsamp \\ r(rtmvt))
 15.       qui svmat randsamp
 16.       qui mvtest normality randsamp1 randsamp2
 17.       mat gof_pvaluesum_mix[`i', `j'] = gof_pvaluesum_mix[`i', `j'] + r(p_dh)
 18.       drop randsamp1 randsamp2
 19.     {\rbr}
 20.     local `j++'
 21.   {\rbr}
 22.   local `i++'
 23. {\rbr}
{\smallskip}
. mat gof_summary_mvn = gof_pvaluesum_mvn/1000
{\smallskip}
. mat gof_summary_mix = gof_pvaluesum_mix/1000

%% file: Example3part1.log.tex
. set seed 3
{\smallskip}
. mat Sigma = J(3, 3, 0.5) + 0.5*I(3)
{\smallskip}
. invmvnormal, p(0.95) mean(0, 0, 0) sigma(Sigma) tail("lower")
quantile = 2.0620839
error = 4.663e-15
flag = 0
fquantile = 0
iterations = 21
{\smallskip}
. local r = r(quantile)

%% file: Example3part2.log.tex
. pmvnormal, lower(., ., .) upper(`r', `r', `r') mean(0, 0, 0) sigma(Sigma)
integral = .94999707
error = .00005633
{\smallskip}
. pmvnormal, lower(., ., .) upper(`= invnormal(1 - 0.05/3)',
> `= invnormal(1 - 0.05/3)', `= invnormal(1 - 0.05/3)') mean(0, 0, 0) sigma(Sigma)
integral = .95705901
error = .00003466

%% file: Example4part1.log.tex
. set seed 4
{\smallskip}
. mat Sigma = J(3, 3, 0.5) + 0.5*I(3)
{\smallskip}
. pmvnormal, lower(0, 0, 0) upper(., ., .) mean(0, 0, 0) sigma(Sigma)
integral = .2500246
error = .00005905

%% file: Example4part2.log.tex
. mat lowertrunc    = J(110, 1, 0)
{\smallskip}
. mat probabilities = J(110, 1, 0)
{\smallskip}
. foreach i of numlist 1/110 {\lbr}
  2.   mat lowertrunc[`i', 1] = -10 + 0.1*(`i' - 1)
  3.   qui tmvnormal, lower(0, 0, 0) upper(., ., .) mean(0, 0, 0) sigma(Sigma) 
> lowert(`=lowertrunc[`i', 1]', `=lowertrunc[`i', 1]', `=lowertrunc[`i', 1]')
> uppert(., ., .)
  4.   mat probabilities[`i', 1] = r(integral)
  5. {\rbr}
{\smallskip}
. mat data = (lowertrunc, probabilities)
{\smallskip}
. qui svmat data
{\smallskip}
. twoway (line data2 data1, yaxis(1)), xtitle(t) ytitle(P(X >= (0,0,0)|(t,t,t)))
> scheme(sj)

%% file: main.bbl
\ifnum 23=1 \def\bibname{Reference}
\else \def\bibname{References} \fi
\begin{thebibliography}{23}
\expandafter\ifx\csname natexlab\endcsname\relax\def\natexlab#1{#1}\fi
\expandafter\ifx\csname url\endcsname\relax
  \def\url#1{\texttt{#1}}\fi
\expandafter\ifx\csname urlprefix\endcsname\relax\def\urlprefix{URL }\fi

\bibitem[{Bailer and Piegorsch(1997)}]{Bailer1997}
Bailer, A.~J., and W.~Piegorsch. 1997.
\newblock \emph{{Statistics for Environmental Biology and Toxicology}}.
\newblock London, United Kingdom: Chapman and Hall.

\bibitem[{Blaesild and Granfeldt(2002)}]{Blaesild2002}
Blaesild, P., and J.~Granfeldt. 2002.
\newblock \emph{{Statistics with Applications in Biology and Geology}}.
\newblock Florida, United States: Chapman and Hall/CRC.

\bibitem[{Cappallari and Jenkins(2006)}]{Cappallari2006}
Cappallari, L., and S.~P. Jenkins. 2006.
\newblock {Calculation of multivariate normal probabilities by simulation, with
  applications to maximum simulated likelihood estimation}.
\newblock \emph{The Stata Journal} 6(2): 156--189.

\bibitem[{Feigelson and Babu(2012)}]{Feigelson2012}
Feigelson, E.~D., and G.~J. Babu. 2012.
\newblock \emph{{Modern Statistical Methods for Astronomy: With R
  Applications}}.
\newblock New York, United States: Cambridge University Press.

\bibitem[{Gates(2006)}]{Gates2006}
Gates, R. 2006.
\newblock {A Mata Geweke-Hajivassiliou-Keane multivariate normal simulator}.
\newblock \emph{The Stata Journal} 6: 190--213.

\bibitem[{Genz(1992)}]{Genz1992}
Genz, A. 1992.
\newblock {Numerical Computation of Multivariate Normal Probabilities}.
\newblock \emph{Journal of Computational and Applied Mathematics} 6(4):
  295--302.

\bibitem[{Genz and Bretz(2002)}]{Genz2002}
Genz, A., and F.~Bretz. 2002.
\newblock {Methods for the Computation of Multivariate t-Probabilities}.
\newblock \emph{Journal of Computational and Graphical Statistics} 11:
  950--971.

\bibitem[{Genz and Bretz(2009)}]{Genz2009}
\mbox{\vrule width30.25006ptheight2.62222ptdepth-2.25222pt}. 2009.
\newblock \emph{{Computation of Multivariate Normal and t Probabilities}}.
\newblock Berlin, Germany: Springer.

\bibitem[{Genz et~al.(2014)Genz, Bretz, Miwa, Mi, Leisch, Scheipl, and
  Hothorn}]{Genz2014}
Genz, A., F.~Bretz, T.~Miwa, X.~Mi, F.~Leisch, F.~Scheipl, and T.~Hothorn.
  2014.
\newblock \emph{{mvtnorm: Multivariate Normal and t Distributions. R package
  version 1.0-0}}.
\urlprefix\url{http://CRAN.R-project.org/package=mvtnorm.}
\bibitem[{Geweke(1989)}]{Geweke1989}
Geweke, J. 1989.
\newblock {Bayesian inference in econometric models using Monte Carlo
  integration}.
\newblock \emph{Econometrica} 66: 1317--1339.

\bibitem[{Gibson et~al.(1994)Gibson, Glasbey, and Elston}]{Gibson1994}
Gibson, G.~J., C.~A. Glasbey, and D.~A. Elston. 1994.
\newblock {Monte Carlo evaluation of multivariate normal integrals and
  sensitivity to variate ordering}.
\newblock In \emph{{Advances in Numerical Methods and Applications}}, ed. I.~T.
  Dimov, B.~Sendov, and P.~S. Vassilevski,  120--126. River Edge: World
  Scientific Publishing.

\bibitem[{Hajivassiliou and McFadden(1998)}]{Hajivassiliou1998}
Hajivassiliou, V., and D.~McFadden. 1998.
\newblock {The method of simulated scores for the estimation of LDV models}.
\newblock \emph{Econometrica} 66: 863--896.

\bibitem[{Howell(2012)}]{Howell2012}
Howell, D.~C. 2012.
\newblock \emph{{Statistical Methods for Psychology}}.
\newblock California, United States: Wadsworth.

\bibitem[{Ireland(2010)}]{Ireland2010}
Ireland, C.~R. 2010.
\newblock \emph{{Experimental Statistics for Agriculture and Horticulture}}.
\newblock Wallingford, United Kingdom: CABI.

\bibitem[{Johnson(1987)}]{Johnson1987}
Johnson, M.~E. 1987.
\newblock \emph{Multivariate Statistical Simulation: A Guide to Selecting and
  Generating Continuous Multivariate Distributions}.
\newblock Chichester, United Kingdom: Wiley.

\bibitem[{Keane(1998)}]{Keane1994}
Keane, M.~P. 1998.
\newblock {A computationally practical simulation estimator for panel data}.
\newblock \emph{Econometrica} 62: 95--116.

\bibitem[{Kotz et~al.(2004)Kotz, Balakrishnan, and Johnson}]{Kotz2004a}
Kotz, S., N.~Balakrishnan, and N.~L. Johnson. 2004.
\newblock \emph{{Continuous Multivariate Distributions, Models and
  Applications}}.
\newblock New York, United States: Wiley.

\bibitem[{Kotz and Nadarajah(2004)}]{Kotz2004b}
Kotz, S., and S.~Nadarajah. 2004.
\newblock \emph{{Multivariate t Distributions and Their Applications}}.
\newblock Cambridge, United Kingdom: Cambridge University Press.

\bibitem[{Mazza and Benaim(2014)}]{Mazza2014}
Mazza, C., and M.~Benaim. 2014.
\newblock \emph{{Stochastic Dynamics for Systems Biology}}.
\newblock Florida, United States: Chapman and Hall/CRC.

\bibitem[{Patel and Read(1996)}]{Patel1996}
Patel, J.~K., and C.~B. Read. 1996.
\newblock \emph{{Handbook of the Normal Distribution}}.
\newblock New York, United States: Marcel Dekker.

\bibitem[{Stevens(2012)}]{Stevens2012}
Stevens, J.~P. 2012.
\newblock \emph{{Applied Multivariate Statistics for the Social Science}}.
\newblock New York, United States: Routledge.

\bibitem[{Tobin(1958)}]{Tobin1958}
Tobin, J. 1958.
\newblock {Estimation of relationships for limited dependent variables}.
\newblock \emph{Econometrica} 26(1): 24--36.

\bibitem[{Tong(2012)}]{Tong2012}
Tong, Y.~L. 2012.
\newblock \emph{{The Multivariate Normal Distribution}}.
\newblock New York, United States: Springer Verlag.

\end{thebibliography}
